\documentclass[preprintnumbers,a4paper,twocolumn,floats,aps,superscriptaddress,nofootinbib,prd]{revtex4-1}

\usepackage{graphicx}
\usepackage{amsmath,amssymb,amsfonts}
\usepackage{verbatim}   
\usepackage{subfigure}  
\usepackage{acronym}
\usepackage{hyperref}
\usepackage{mathrsfs}
\usepackage{multirow}
 \usepackage{slashed}
 \usepackage{url}
 \usepackage{color}

\usepackage{amsmath}
\usepackage{amsfonts}
\usepackage{amssymb}

\begin{document}

\title{Determining the structure of dark-matter couplings  at the LHC}

\preprint{OUTP-13-23P}

\author{Ulrich Haisch}
\email{u.haisch1@physics.ox.ac.uk}
\affiliation{Rudolf Peierls Centre for Theoretical Physics, University of Oxford, 1 Keble Road, Oxford OX1 3NP, United Kingdom}
\author{Anthony Hibbs}
\email{Anthony.Hibbs@physics.ox.ac.uk}
\affiliation{Rudolf Peierls Centre for Theoretical Physics, University of Oxford, 1 Keble Road, Oxford OX1 3NP, United Kingdom}
\author{Emanuele Re}
\email{Emanuele.Re@physics.ox.ac.uk}
\affiliation{Rudolf Peierls Centre for Theoretical Physics, University of Oxford, 1 Keble Road, Oxford OX1 3NP, United Kingdom}

\begin{abstract} 
The latest LHC mono-jet searches place stringent  bounds on the $pp \to \bar \chi \chi$  cross section of dark matter. Further properties such as the dark matter mass or the precise structure of the interactions between dark matter and the standard model can however not be determined in this manner. We point out that measurements of the azimuthal angle correlations between the two jets in $2 j + \bar \chi \chi$ events may be used to disentangle whether dark matter pair production proceeds dominantly through tree or loop diagrams. Our general observation is illustrated by considering theories in which dark matter interacts predominantly with the top quark. We  show explicitly  that in this case the jet-jet azimuthal angle difference is a gold-plated observable to probe the Lorentz structure of the couplings of dark matter to top quarks, thus testing the CP nature of the particle mediating these interactions. 
\end{abstract}

\maketitle

\section{Introduction}
\label{sec:introduction}

The minimal experimental signature of dark matter~(DM) pair production at the LHC would be an excess of events with a single jet in association with large amounts of missing transverse energy ($E_{T, {\rm miss}}$). The experimental search for $j + E_{T, {\rm miss}}$ events provides  bounds on the interaction strength of DM with quarks and gluons, constraining the same parameters as direct detection experiments (see~e.g.~\cite{ATLAS,CMS}). These measurements place the leading (and in some cases only) limits on models of DM over certain regions of parameter space. 

While the  $j + E_{T, {\rm miss}}$  channel  can be used to constrain the $\sigma (pp \to \bar \chi \chi)$   cross section, it provides insufficient information to determine additional DM properties such as its mass or the precise nature of its interactions with the standard model (SM). In fact,  the transverse momentum~($p_T$) spectrum of the $j + E_{T, {\rm miss}}$  signal is essentially featureless and almost independent of the chirality and/or the CP properties of the DM couplings to quarks.\footnote{For instance, the $p_T$ spectra corresponding to effective vector and axial-vector DM-quark interactions are within the uncertainties present at the next-to-leading order (NLO) plus parton-shower~(PS) level \cite{Haisch:2013ata} indistinguishable.} This suggests that while ATLAS and CMS are well suited to discover light DM, the LHC prospects of using this channel to make more definitive statements about specific DM properties  seem to be slim. 

\begin{figure}[!t]
\includegraphics[height=0.175\textwidth]{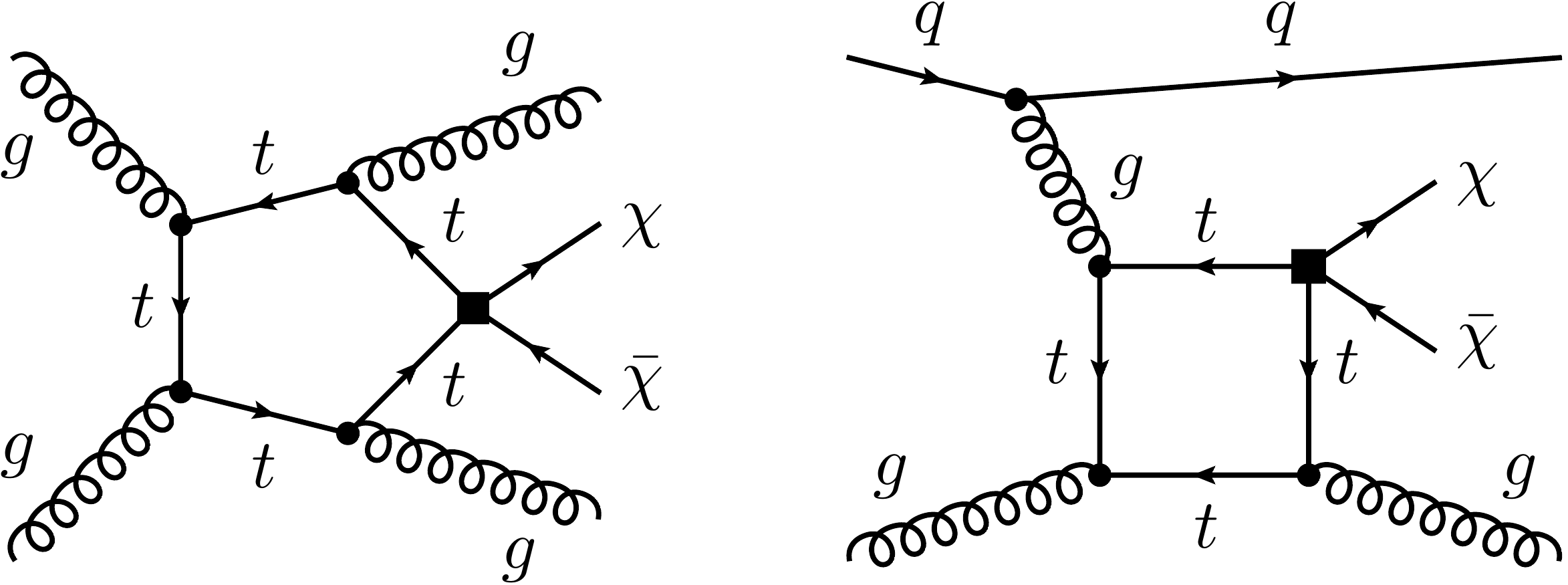} 
\vspace{0mm}
\caption{Typical one-loop diagrams leading to  $2 j + E_{T,\rm miss}$ events in $pp$ collisions. The black squares denote insertions of the four-fermion operators ${\cal O}_{S,P}$.}
\label{Fig1}
\end{figure}

In this letter we observe that this  unsatisfactory situation may be remedied by studying two-jet final states involving $E_{T, {\rm miss}}$. In particular, we will argue that measurements of the azimuthal angle difference   in $2 j + E_{T, \rm miss}$ events can possibly show a strong cosine-like or sine-like correlation only if DM pair production is loop induced, whereas tree-level interactions result in a  $\Delta \phi_{j_1j_2}$ distribution of a quite different shape. In order to illustrate our general observation we will consider DM models that generate the effective operators 
\begin{equation} \label{eq:OSP}
\mathcal{O}_S = \frac{m_t}{\Lambda^3_S} \, \bar{t} t \, \bar{\chi} \chi \, , \qquad 
\mathcal{O}_P = \frac{m_t}{\Lambda^3_P} \, \bar{t} \gamma_5 t \, \bar{\chi} \gamma_ 5 \chi \, .
\end{equation}
Examples of Feynman diagrams with an insertion of ${\cal O}_{S,P}$ that give rise to a $2j + E_{T, {\rm miss}}$ signal are displayed in~Fig.~\ref{Fig1}. For this well-motivated case we will explicitly show that the Lorentz structure of the DM top-quark interactions --- and in consequence the CP nature of the mediator inducing (\ref{eq:OSP})  --- can be disentangled by measuring the normalised $\Delta \phi_{j_1j_2}$ distribution. After a discovery of an enhanced mono-jet signal, combining the measurements of the top-loop induced $\sigma (pp \to j + \bar \chi \chi)$  cross section \cite{Haisch:2012kf,Fox:2012ru} and the $1/\sigma \, d\sigma (pp \to 2 j + \bar \chi \chi)/d \Delta \phi_{j_1j_2}$  spectrum of the jet-jet azimuthal angle difference would hence not only allow to  determine the suppression scales $\Lambda_{S,P}$ in (\ref{eq:OSP}) but also  whether the scalar operator~${\cal O}_S$ or the pseudo-scalar operator~${\cal O}_P$ is responsible for the observed excess of $j+E_{T, {\rm miss}}$ events. Other constraints on effective interactions between DM and top quarks have been discussed for example in \cite{Haisch:2013uaa,Lin:2013sca}. 

Our work is organised as follows: in Sec.~\ref{sec:interactions} we introduce the DM interactions which we intend to examine. In~Sec.~\ref{sec:2JetsDM} we calculate the azimuthal angle correlations of the two jets in $2 j + \bar \chi \chi$ production induced by the operators ${\cal O}_{S,P}$, including the full top-quark mass dependence of the squared matrix elements. Our calculation is performed at the leading order (LO) in QCD. We will also comment on the applicability of the heavy top-quark approximation and the impact of higher-order QCD effects. In Sec.~\ref{sec:discussion} we discuss the case where the mediator can be resonantly produced, before concluding in Sec.~\ref{sec:conclusions}. 

\section{DM interactions}
\label{sec:interactions}

In the following we are interested in DM pair production from quark or gluon initial states. We will restrict our discussion to the case where the production proceeds via the exchange of a spin-$0$  $s$-channel mediator. We consider the following interactions between DM and top quarks involving a colourless scalar ($S$) or pseudo-scalar~ ($P$) mediator\footnote{LHC constraints on the scalar and pseudo-scalar DM-quark interactions involving the light flavours $q = u,d,s,c,b$ have been discussed in \cite{Haisch:2013ata,Fox:2012ru}.}
\begin{equation} \label{eq:LSP}
\begin{split}
\mathcal{L}_S   & = g_\chi^S \left(  \bar \chi   \chi  \right ) S +  g^S_t \, \frac{m_t}{v} \left(  \bar t t  \right ) S   \,, \\[2mm]
\mathcal{L}_P   & =  i g_\chi^P \left(  \bar \chi  \gamma_5 \chi  \right ) P + i g^P_t \, \frac{m_t}{v}  \left(  \bar t   \gamma_5 t  \right )P \,,
\end{split}
\end{equation}
where $v \simeq 246 \, {\rm GeV}$ is the Higgs vacuum expectation value. Notice that we have assumed  that the couplings of the mediators to top quarks are proportional to the associated SM Yukawa coupling. This is motivated by the hypothesis of minimal flavour violation (MFV), which curbs the size of dangerous flavour-changing neutral current processes and  automatically leads to a stable DM candidate~\cite{Batell:2011tc}. While the DM particle $\chi$ in (\ref{eq:LSP}) is understood to be a Dirac fermion, extending our discussion to Majorana DM or the case of a complex/real scalar is straightforward~(see~\cite{Haisch:2012kf} for details).

If the mediator masses $M_{S,P}$ are large  compared to the invariant mass $m_{\bar \chi \chi}$ of the DM pair, we can describe $2 j + \bar \chi \chi$ production by means of an effective field theory (EFT). Integrating out the scalar and pseudo-scalar mediator then gives rise to~(\ref{eq:OSP}) as well as  composite operators consisting of four top-quark fields, which we do not consider further.\footnote{Unlike the operator $\bar t t \, \bar t \gamma_5 t$, which is strongly constrained because it contributes to the electric dipole moment of the neutron  \cite{Brod:2013cka}, the purely scalar or pseudo-scalar four-top operators resulting from~(\ref{eq:LSP}) are experimentally not well bounded. The appearance of the operator $\bar t t \, \bar t \gamma_5 t$ can be avoided by taking the spin-0 mediators $S, P$ to be CP eigenstates.} In the case of ${\cal O}_S$ the suppression scale~$\Lambda_S$ is related to the mediator mass and the fundamental couplings by
\begin{equation} \label{eq:Lambda}
\Lambda_S = \left ( \frac{v M_S^2}{g_\chi^S g_t^S} \right )^{1/3} \,,
\end{equation}
and an analogous expression with $S \to P$ holds for ${\cal O}_P$. 

With the current $j + E_{T,\rm miss}$ \cite{Haisch:2012kf} and $\bar t t + E_{T,\rm miss}$ \cite{Lin:2013sca} data, one can exclude values of the suppression scale below roughly $150 \, {\rm GeV}$ ($170 \, {\rm GeV}$) in the scalar (pseudo-scalar) case for light DM, which is small compared to typical LHC energies. In order to discuss the validity of the EFT approach (see also \cite{Fox:2011pm,Shoemaker:2011vi,Fox:2012ee,Busoni:2013lha,Profumo:2013hqa,Buchmueller:2013dya}), we  will consider in Sec.~\ref{sec:discussion} also the simplest ultraviolet (UV) completion, where (\ref{eq:OSP}) arises from the full theory (\ref{eq:LSP}) after integrating out the  fields $S$ and $P$. We will see that in this case the analysis becomes more model-dependent, because the predictions now depend on $g^{S,P}_t$ and $g^{S,P}_\chi$ as well as the masses $M_{S,P}$ and the decay widths $\Gamma_{S,P}$ of the mediators. Apart from these minor complications our general conclusions will however also hold in the case where the $s$-channel resonances $S,P$ can be directly produced in $pp$ collisions. 

\section{DM production with two jets}
\label{sec:2JetsDM}

In our analysis we consider $2 j+ E_{T,\rm miss}$ production at the LHC with $\sqrt{s} = 14 \, {\rm TeV}$ center-of-mass (CM) energy. We adopt event selection criteria corresponding to the latest  CMS  mono-jet search~\cite{CMS}.\footnote{The cuts imposed in the existing ATLAS and CMS analyses will not be suitable for DM searches at the $14 \, {\rm TeV}$ LHC due to triggering limitations \cite{Schramm}.  Our work should hence only be considered as a proof of concept. A more realistic study, including NLO corrections, PS effects and hadronisation corrections for both the DM signal and the SM backgrounds, will be presented elsewhere~\cite{inpreparation}.}  In this search events of more than two jets with  pseudo-rapidity below 4.5 and transverse momentum above $30 \, {\rm GeV}$ are rejected. In order to suppress QCD di-jet events, CMS puts an angular requirement on the azimuthal distance between the two tagging jets of $\Delta\phi_{j_1j_2} < 2.5$. Our  reference signal region is defined by $|\eta_{j_1}|<2.4$, $p_{T,j_1}>110 \, {\rm GeV}$ and $E_{T,{\rm miss}}> 350 \, {\rm GeV}$, but we will comment on the sensitivity of the  signal on  the $p_{T,j_1}$ and  $E_{T,{\rm miss}}$ cuts. To improve the separation between the azimuthal angle distribution of the SM background and the $2 j+ E_{T,\rm miss}$ signal, we also impose a cut of $m_{j_1 j_2} > 600 \, {\rm GeV}$ on the invariant mass of the di-jet system.

The calculation of the azimuthal distance $\Delta \phi_{j_1j_2}$  of the $2 j + E_{T,\rm miss}$ signal events is performed with the help of {\tt GGFLO} which is part of {\tt VBFNLO} \cite{Arnold:2012xn}, modifying the process $pp \to 2j+ h \, (A)$ appropriately. The  {\tt GGFLO} implementation of the $2j+ h \, (A)$ production process is based on the analytical LO results of~\cite{DelDuca:2001eu,DelDuca:2001fn} for the scalar Higgs~($h$) case and of \cite{Campanario:2010mi} for the pseudo-scalar Higgs~($A$) case. Our simulations utilise MSTW2008LO parton distributions~\cite{Martin:2009iq} and jets are constructed according to the anti-$k_t$ algorithm \cite{Cacciari:2008gp} with a radius parameter of $R=0.5$, which corresponds to the value used in the CMS analysis~\cite{CMS}. 

\begin{figure}[!t]
\includegraphics[width=0.475\textwidth]{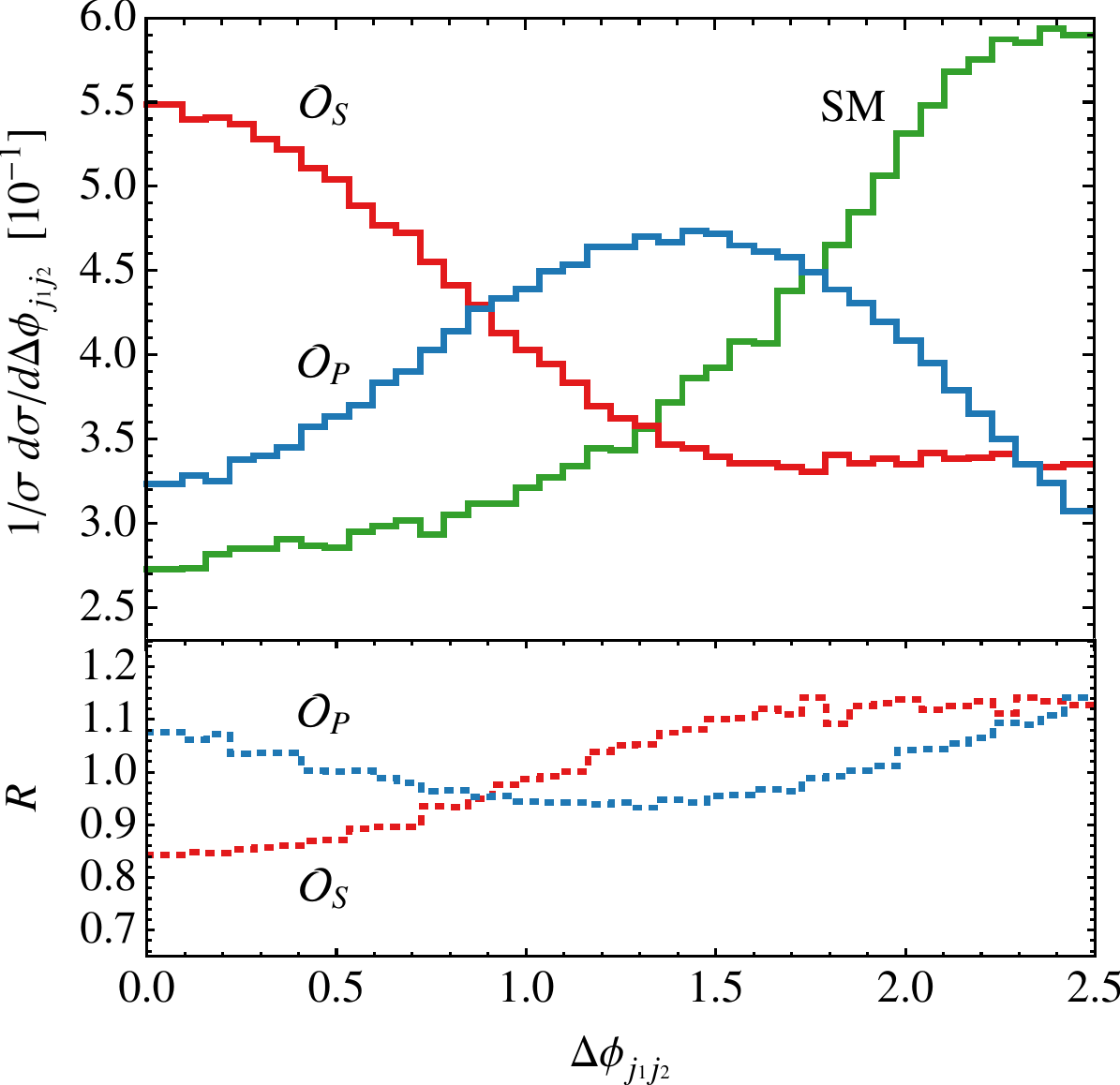}
\vspace{0mm}
\caption{Normalised $\Delta \phi_{j_1j_2}$ distribution for the insertion of ${\cal O}_S$~(red)  and ${\cal O}_P$~(blue) applying the cuts $p_{T,j_1} > 110 \, {\rm GeV}$ and $E_{T, \rm miss} > 350 \, {\rm GeV}$. The solid curves correspond to the full results while the dotted curves show the ratio $R$ between the results in the heavy top-quark mass approximation and the exact predictions. For comparison the green solid curve indicates the prediction of the dominant SM background process, $pp \to 2j + Z (\to  \bar \nu \nu)$, using the same event selection criteria. For further details see text.}
\label{Fig2}
\end{figure}

We start our numerical analysis by showing results  obtained for $\Lambda_{S,P} = 150 \, {\rm GeV}$, a DM mass of $m_\chi = 50 \, {\rm GeV}$, employing the reference cuts described above.  Our  choice of parameters will be motivated in Sec.~\ref{sec:discussion}. The central values of the corresponding $j+E_{T, \rm miss}$ and $2j+E_{T, \rm miss}$ signal cross sections are  $675 \, {\rm fb}$ and $204 \, {\rm fb}$ ($1119 \, {\rm fb}$ and $338 \, {\rm fb}$) for ${\cal O}_S$~(${\cal O}_P$), while the SM background predictions amount to  $1289 \, {\rm fb}$ and $330 \, {\rm fb}$. To put these numbers into perspective we recall that the latest CMS analysis~\cite{CMS} excludes excesses in the mono-jet cross section with  signal-over-background ratios of $S/B \gtrsim 0.15$ at 95\%~confidence level.  Given these numbers the mono-jet  signals corresponding to $\Lambda_{S,P} = 150 \, {\rm GeV}$ and $m_\chi = 50 \, {\rm GeV}$ should  be easily detectable at the $14 \, {\rm TeV}$ LHC. 

The normalised $\Delta \phi_{j_1j_2}$ distributions associated to the operators ${\cal O}_{S,P}$ are displayed in~Fig.~\ref{Fig2}. From the figure it is evident that the scalar operator ${\cal O}_S$ produces a strong correlation between the two jets, with a  distribution that is peaked at $\Delta \phi_{j_1 j_2}  =0$ and heavily suppressed at  $\Delta \phi_{j_1 j_2} = \pi/2$~(red solid curve). In the case of the pseudo-scalar operator~${\cal O}_P$ the position of the peak and trough is instead reversed~(blue solid curve). The cosine-like (sine-like) modulation in the azimuthal angle distribution corresponding to ${\cal O}_S$~(${\cal O}_P$) should be contrasted with the spectrum of the dominant SM background process,  $pp \to 2j + Z (\to  \bar \nu \nu)$, which has a minimum at $\Delta \phi_{j_1 j_2} = 0$ and a maximum in the vicinity of $\Delta \phi_{j_1 j_2} = 2.5$~(green solid curve). We simulate the background at LO using the {\tt POWHEG~BOX}~\cite{Alioli:2010xd,Re:2012zi}. PS effects or hadronisation  corrections are not included in our SM prediction.  

To assess the significance of our findings,  we study the scale uncertainties of the results. As advocated in \cite{DelDuca:2001fn}, we identify the factorisation scale as $\mu_F = \xi \hspace{0.5mm} (p_{T, j_1} \hspace{0.25mm} p_{T, j_2})^{1/2}$ and replace the overall factor $\alpha_s^4$ entering the $2 j + E_{T, \rm miss}$ cross section by $\alpha_s (\xi \hspace{0.25mm} p_{T, j_1}) \hspace{0.25mm} \alpha_s (\xi \hspace{0.25mm} p_{T, j_2})  \hspace{0.25mm} \alpha_s^2 (\xi \hspace{0.25mm}  m_{\bar \chi \chi})$. We evaluate these quantities for every event generated by our Monte Carlo (MC) and vary $\xi$ in the range $[1/2,2]$. In the total cross sections the induced scale uncertainties are around $^{+80\%}_{-40\%}$, while the relative shifts in the normalised differential azimuthal angle distributions do not exceed the level of $^{+  5 \%}_{-  5 \%}$.  We conclude from this that even after considering scale ambiguities, the normalised $\Delta \phi_{j_1j_2}$ distribution for~${\cal O}_S$ is different than that of ${\cal O}_P$, and both spectra are clearly distinguishable from the SM background.
 
The distinction in the radiation pattern of ${\cal O}_S$ and ${\cal O}_P$ can be most easily understood by employing the heavy top-quark mass limit. In fact, in this approximation the effect of top-quark loops in $2 j + E_{T,\rm miss}$ production can be described  in terms of the following two effective operators
\begin{equation} \label{eq:OGG}
\begin{split}
{\cal O}_G & = \frac{\alpha_s}{12 \pi \hspace{0.25mm} \Lambda_S^3} \, G_{\mu\nu}^a G^{a, \mu \nu} \, \bar \chi \chi \,, \\[2mm]
{\cal O}_{\widetilde G} & = \frac{\alpha_s}{8 \pi \hspace{0.25mm} \Lambda_P^3} \, G_{\mu\nu}^a \widetilde G^{a, \mu \nu} \, \bar \chi \gamma_5 \chi \,, 
\end{split}
\end{equation}
where $G_{\mu \nu}^a$ denotes the gluon field strength tensor and $\widetilde G^{a, \mu \nu} = 1/2 \hspace{0.5mm} \epsilon^{\mu\nu\lambda\rho} \hspace{0.25mm} G_{\lambda\rho}^a$ its dual. 

In the limit that the external partons only experience a small energy loss and that the momentum components of the tagging jets in the beam direction are much greater than those in the transverse plane, the structure of the $pp \to 2j + \bar \chi \chi$ matrix element of ${\cal O}_G$ and  ${\cal O}_{\widetilde G}$ is easy to work out \cite{Plehn:2001nj}. Denoting the currents and momenta of the gluons that initiate the scattering by  $J_{1,2}$ and $q_{1,2}$, one finds in the case of ${\cal O}_G$  the result ${\cal M}_G \sim   J_1^\mu J_2^\nu \hspace{0.5mm} (g_{\mu\nu} \hspace{0.25mm} q_1 \cdot q_2 - q_{1 \nu} q_{2 \mu}) \sim \vec{p}_{T,j_1} \cdot \vec{p}_{T,j_2}$. This implies that the $\Delta \phi_{j_1 j_2}$ spectrum corresponding to ${\cal O}_S$ should be enhanced for collinear tagging jets, $\Delta \phi_{j_1 j_2} =0$, while for  $\Delta \phi_{j_1 j_2} =\pi/2$ it should show  an approximate zero. In the case of  ${\cal O}_{\widetilde G}$ one obtains instead ${\cal M}_{\widetilde G} \sim   \epsilon_{\mu\nu\lambda\rho} \hspace{0.25mm} J_1^\mu J_2^\nu q_1^\lambda \hspace{0.25mm} q_2^\rho \sim \vec{p}_{T,j_1} \times \vec{p}_{T,j_2}$. It follows that the  $\Delta \phi_{j_1 j_2} $ distribution for ${\cal O}_P$ should have a dip if the two jets are  collinear, $\Delta \phi_{j_1 j_2} =0$, or back-to-back, $\Delta \phi_{j_1 j_2} = \pi$, as the Levi-Civita tensor forces the result to zero. These features are clearly visible in~Fig.~\ref{Fig2}. The above discussion also implies that  in any theory in which one of the loop-induced  operators in (\ref{eq:OGG}) is generated, the azimuthal angle difference   in $2 j + E_{T, \rm miss}$ events will show a strong cosine-like or sine-like correlation. In theories in which DM  pair production proceeds dominantly via tree-level graphs this will not be the case. Measurements of the $\Delta \phi_{j_1 j_2}$ spectrum are thus in principle sensitive to the quantum structure of the DM interactions with the SM. 

The lower part of Fig.~\ref{Fig2} also shows that while the predictions obtained in the heavy top-quark mass approximation  (dotted red and blue curves) describe the full results (solid red and blue curves) within an accuracy of  $\pm 20\%$ or better, taking this limit always reduces the amplitude of the cosine-like and sine-like modulations. The behaviour found for $1/\sigma \, d\sigma (pp \to 2 j + E_{T, \rm miss})/d\Delta \phi_{j_1 j_2}$ is in clear contrast to that obtained in the case of the loop-induced mono-jet cross section for which the limit $m_t \to \infty$ is not a good approximation~\cite{Haisch:2012kf}, because the high-$p_T$ jet is able to resolve the sub-structure of the top-quark loop.  In fact, also in the case of $\sigma( pp \to 2 j + \bar \chi \chi)$, we find that the  EFT predictions and the exact results are vastly different. For our standard cuts the infinite top-quark mass approximation overestimates the $2 j + E_{T, \rm miss}$ cross section by a factor of around $7$~($10$) in the case of the operator ${\cal O}_S$~(${\cal O}_P$). 

\begin{figure}[!t]
\includegraphics[width=0.475\textwidth]{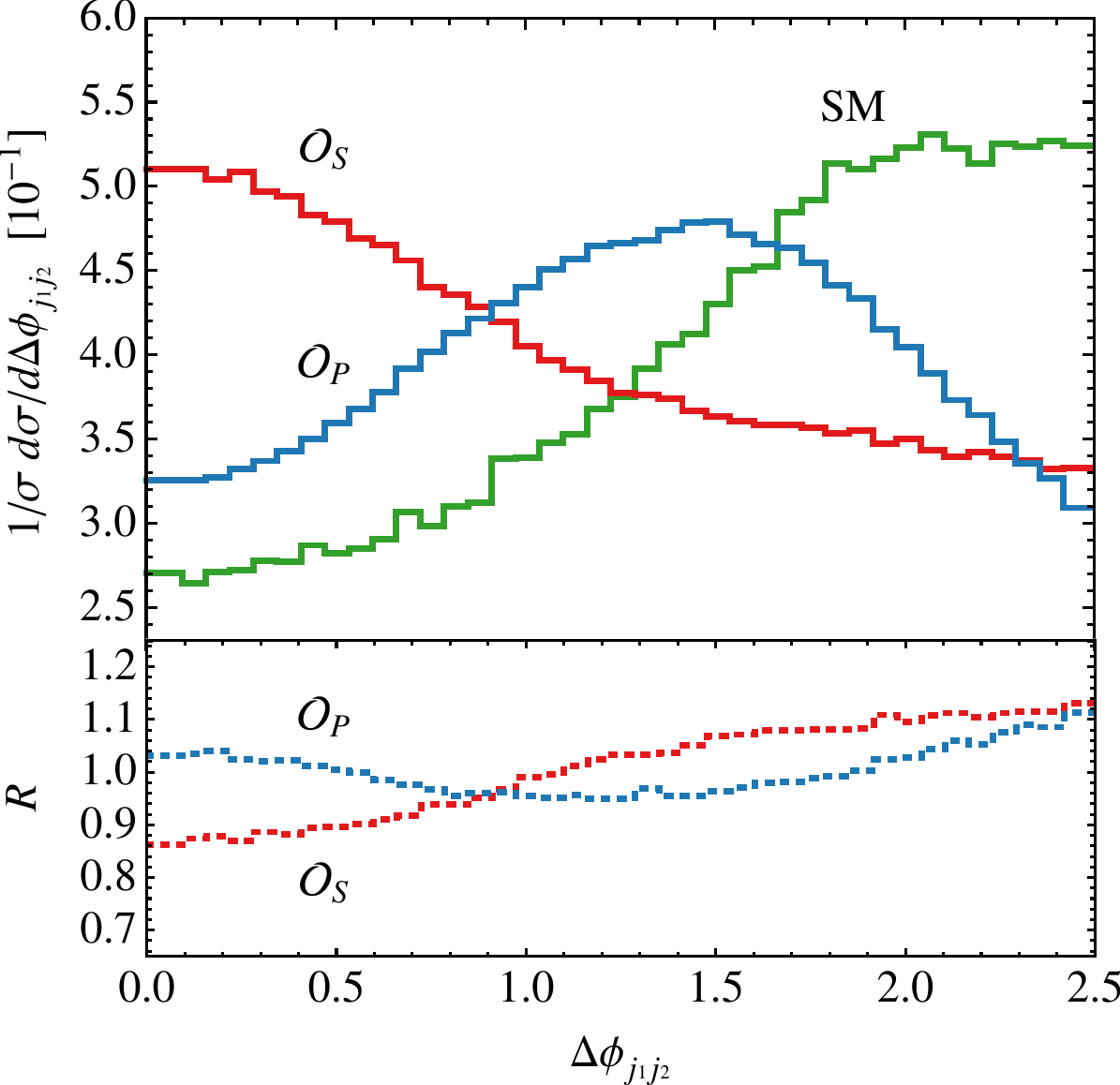}
\vspace{0mm}
\caption{Normalised $\Delta \phi_{j_1j_2}$ distributions using the event selection criteria $p_{T,j_1} > 350 \, {\rm GeV}$ and $E_{T, \rm miss} > 500 \, {\rm GeV}$.
The style and colour coding of the curves follows the one used in~Fig.~\ref{Fig2}.}
\label{Fig3}
\end{figure}

In order to further illustrate this point we show in~Fig.~\ref{Fig3} the normalised $\Delta \phi_{j_1 j_2}$ distributions for ${\cal O}_{S,P}$ using again $\Lambda_{S,P} = 150 \, {\rm GeV}$ and  $m_\chi = 50 \, {\rm GeV}$, but applying the stronger  signal cuts $p_{T,j_1} > 350 \, {\rm GeV}$ and $E_{T, \rm miss} > 500 \, {\rm GeV}$.  The corresponding  $j+E_{T, \rm miss}$ and $2j+E_{T, \rm miss}$ cross sections read  $214 \, {\rm fb}$ and $87 \, {\rm fb}$ (${\cal O}_S$), $344 \, {\rm fb}$ and $141 \, {\rm fb}$ (${\cal O}_P$) and   $246 \, {\rm fb}$ and  $92 \, {\rm fb}$ (SM). One first observes that the infinite top-quark mass limit still furnishes an acceptable description of the full results in this case. Second, the cosine-like and sine-like modulations of the $\Delta \phi_{j_1 j_2}$ spectra are less  pronounced if the requirements on  $p_{T,j_1}$ and $E_{T, \rm miss}$ are more exclusive. This feature can be understood by recalling that a pure cosine-like or sine-like $\Delta \phi_{j_1 j_2}$ spectrum requires that the transverse momenta of the jets are much smaller than the momentum components  along the beam direction. For harder $p_{T, j_1}$ cuts this approximation is not as good and as a result the strong jet-jet correlation is less marked. We conclude from this that in order to maximise the power of the $\Delta \phi_{j_1 j_2}$ distribution in determining the Lorentz structure of the DM top-quark interactions the $p_{T,j_1}$ and $E_{T, \rm miss}$ cuts should be as loose as possible. Making this statement more precise would require to perform a dedicated analysis of the cut dependencies of both the signal and the background. While such a study is beyond the scope of this letter, we plan to return to this question in a future publication \cite{inpreparation}. 

Another important and related issue is the question whether higher-order QCD effects can potentially wash out the observed strong correlations between the two jets. This question can be addressed by relying again on the similarities of the signal process $pp \to 2 j + E_{T, \rm miss}$ and its QCD analog $pp \to 2 j + h \, (A)$. In the latter case it has been shown by explicit calculations (see~e.g.~\cite{DelDuca:2006hk,Campbell:2006xx,Andersen:2010zx}) that the shape of the lowest order distributions are unchanged and that therefore the jet-jet correlations survive the addition of NLO QCD corrections as well as PS and hadronisation effects.  We verified that the latter feature is also present in the case of $1/\sigma \, d\sigma (pp \to 2 j + \bar \chi \chi)/d \Delta \phi_{j_1j_2}$  by showering our LO results with  PYTHIA~6.4~\cite{Sjostrand:2006za}. We find that PS effects result in relative shifts of maximal $^{+8\%}_{-8\%}$ in the $\Delta \phi_{j_1 j_2}$ distributions and slightly reduce the amplitudes of the cosine-like and sine-like modulations, but do not distort the spectra. Given its stability under radiative corrections, we believe that the normalised spectrum of the  azimuthal angle difference $\Delta \phi_{j_1j_2}$ in $2 j + \bar \chi \chi$ production is a gold-plated observable for determining  the structure of the couplings of DM to top quarks.  

\section{Discussion}
\label{sec:discussion}

Until now we have considered an EFT framework to interpret a hypothetical mono-jet  signal. This is particularly simple because in such a case the complete information is encoded in the scales $\Lambda_{S,P}$ that suppress the effective couplings (\ref{eq:OSP}), making it unnecessary to specify details of the particle mediating the interactions. Given the weakness of the bounds on $\Lambda_{S,P}$  \cite{Haisch:2012kf,Lin:2013sca}, there are however serious concerns regarding the validity of the EFT approach (see also \cite{Fox:2011pm,Shoemaker:2011vi,Fox:2012ee,Busoni:2013lha,Profumo:2013hqa,Buchmueller:2013dya} for similar discussions). In this section, we will therefore quantify when the simple-minded limits on the scale of the  scalar and pseudo-scalar interactions apply and under which  circumstances the EFT framework breaks down. In order to go beyond the effective description, one has to specify a concrete UV completion. In the following, we will assume that the full theory is provided by (\ref{eq:LSP}), which implies that the effective interactions (\ref{eq:OSP}) are generated by the $s$-channel exchange of the colourless spin-0 states $S,P$. We will not discuss the case of $t$-channel exchange of coloured  spin-0  mediators, which is interesting in its own right and has been utilised in~\cite{Kumar:2013hfa, Batell:2013zwa} to construct MFV DM models where the relic carries  top flavour. 

We follow \cite{Buchmueller:2013dya} to determine the  minimum value of the couplings $(g_\chi^{S,P} g_t^{S,P})^{1/2} = (v M_{P,S}^2/\Lambda_{P,S}^3)^{1/2}$ for which the EFT approach is applicable. First, we derive the limits on the suppression scales $\Lambda_{S,P}$ as a function of the DM mass $m_\chi$. For concreteness, our analysis is based on the most recent mono-jet search by CMS~\cite{CMS} with an integrated luminosity of $19.5 \, {\rm fb}^{-1}$ at $\sqrt{s} = 8 \, {\rm TeV}$, utilising our standard event selection criteria. Second, we calculate $\sigma (pp \to j + E_{T, \rm miss})$ in the full theory as a function of both $m_\chi$ and $M_{S,P}$. The actual computation of the top-loop induced $j + E_{T, \rm miss}$ cross sections is performed by means of the MC codes developed in~\cite{Haisch:2012kf,Fox:2012ru}, which give identical results. For each DM mass, the  minimum value of  $(g_\chi^{S,P} g_t^{S,P})^{1/2}$ consistent with an EFT description is then found from the requirement that the full theory calculation of $\sigma (pp \to j + E_{T, \rm miss})$ agrees with the corresponding EFT result to better than $20\%$.  In the whole procedure, we take into account that   $\Lambda_{S,P}$ and $M_{S,P}$ are related via (\ref{eq:Lambda}). 

\begin{figure}[!t]
\includegraphics[height=0.45\textwidth]{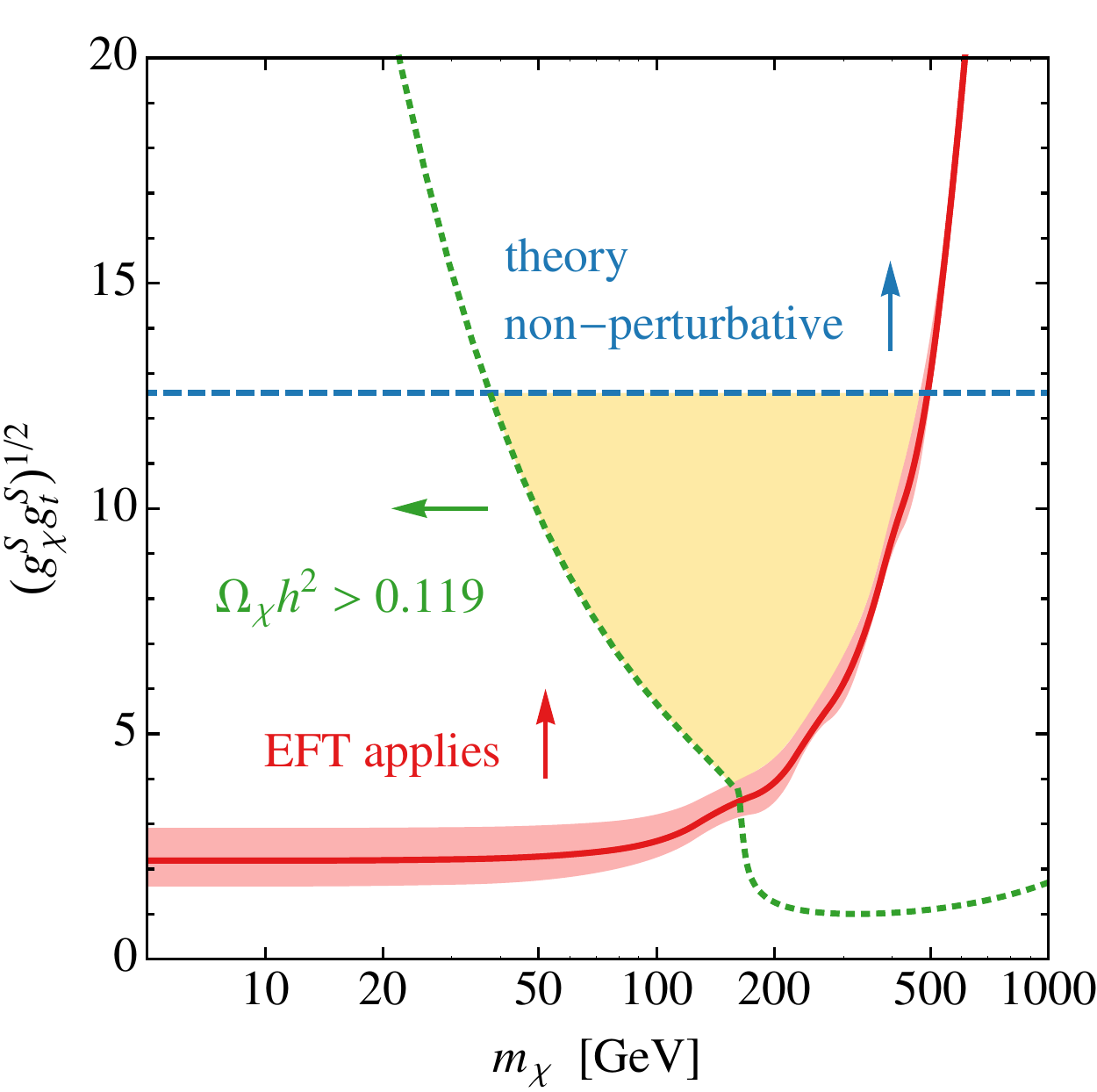} 

\includegraphics[height=0.45\textwidth]{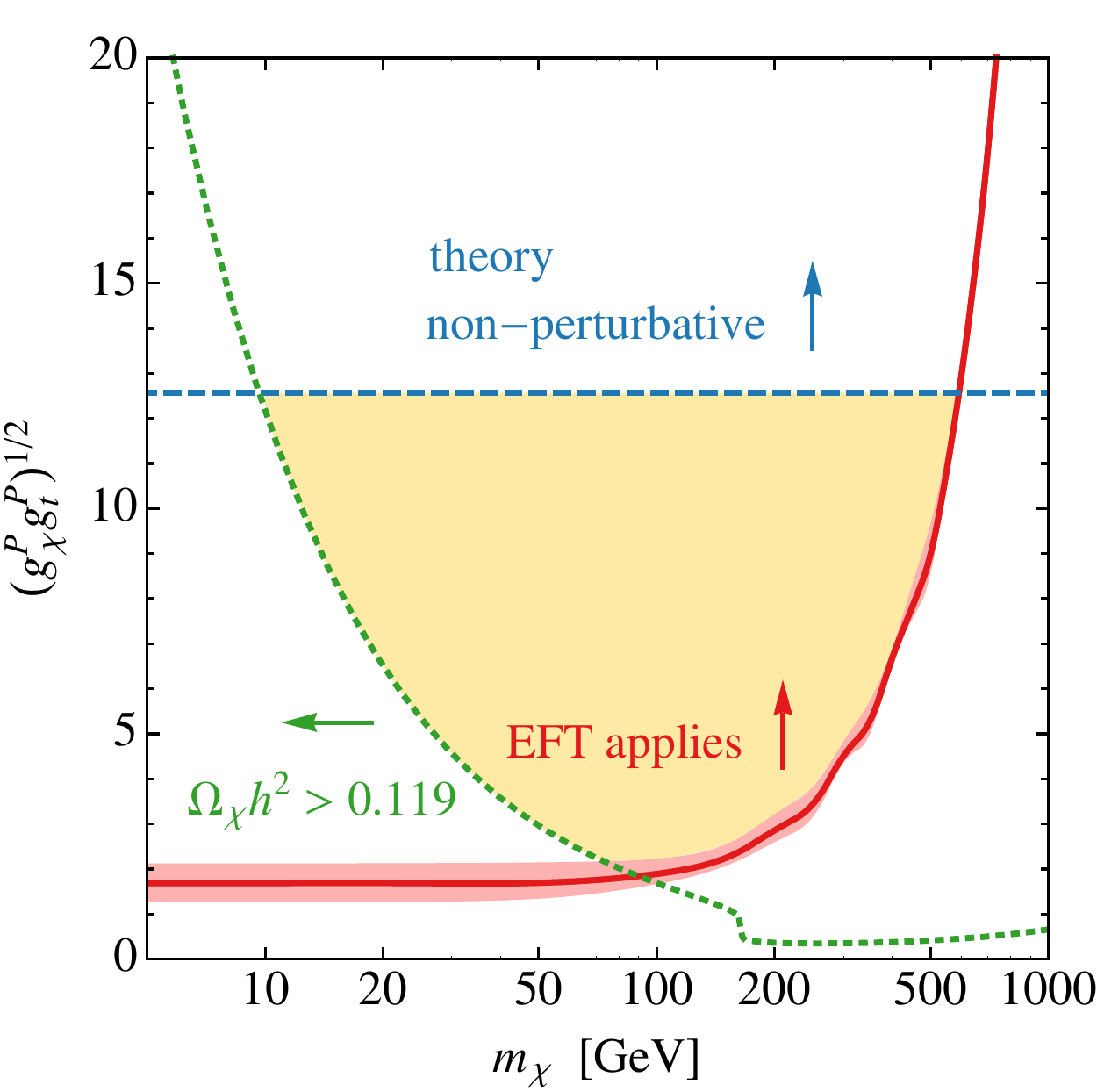} 
\vspace{2mm}
\caption{Upper panel: The red solid curve and band indicates the minimum value of $(g_\chi^S g_t^S)^{1/2}$ for which the LHC bounds on $\Lambda_S$ hold. The perturbative limit on this combination of couplings  is indicated by the  blue dashed curve, while the green dotted curve marks  the parameter space where the DM relic density agrees with observation. Lower panel: The analogous bounds on $(g_\chi^P g_t^P)^{1/2}$. In both panels the region of parameter space compatible with all constraints is coloured yellow. See text for further explanations.}
\label{Fig4}
\end{figure}

The minimal coupling strengths determined in this manner are indicated by the red solid curves and bands in~Fig.~\ref{Fig4}. The width of the bands reflects the dependence of the predictions on the relative width of the mediators, which we vary in the range $\Gamma_{S,P}/M_{S,P} \in [ 1/(8 \pi), 1/3]$  to obtain the shown results. We see that for the EFT to work the couplings of the $s$-channel mediators to DM and top quarks have to be strong and that increasingly  larger values of  $(g_\chi^{S,P} g_t^{S,P})^{1/2}$ are needed for an accurate description, if the DM mass lies at or above the weak scale. In fact, in the case of ${\cal O}_S$~(${\cal O}_P$) the theory becomes necessarily non-perturbative for  $m_\chi \gtrsim  490 \, {\rm GeV}$ ($m_\chi \gtrsim 580 \, {\rm GeV}$) as indicated by the blue dashed curves in the plots. It is important to realise that the values $M_{S,P}$ for which the EFT is applicable are  below a TeV if DM is light. To give an example, for $m_\chi = 50 \, {\rm GeV}$ the displayed EFT limits correspond to $M_S \simeq 370 \, {\rm GeV}$ and $M_P \simeq 310 \, {\rm GeV}$, respectively, if one assumes that the relative widths are $\Gamma_{S,P}/M_{S,P} = 1/3$. 

The DM relic abundance  also depends on the couplings $g_{\chi,t}^{S,P}$ and the masses $M_{S,P}$. However, this observable is sensitive to the full particle content of the underlying UV theory, because the mass spectrum determines the number and the strengths of the DM annihilation channels. This feature makes the prediction for $\Omega_\chi h^2$  more model-dependent than the mono-jet cross sections analysed above. For simplicity, we will assume that the couplings and the particle content are completely specified by (\ref{eq:LSP}), meaning that only annihilation processes with top quarks and gluon pairs in the final state are possible. We also allow for either scalar or pseudo-scalar interactions but not both. 

Using the relevant formulas for the annihilation cross sections given in \cite{Haisch:2012kf} and requiring that  the relic abundance saturates the observed value $\Omega_\chi h^2 = 0.119$~\cite{Ade:2013zuv}, we find the green dotted curves in the panels of~Fig.~\ref{Fig4}. The parameter regions to the left  and right of the curves correspond to DM overproduction and underproduction in the early universe. From the intersections of the non-perturbativity bounds and  the relic density constraints, we obtain the following  limit $m_\chi \gtrsim  40 \, {\rm GeV}$ ($m_\chi \gtrsim 10 \, {\rm GeV}$)  in the case of the operator ${\cal O}_S$~(${\cal O}_P$). Combining all constraints we then find  the yellow coloured wedges, which correspond to strongly-coupled theories with weak scale DM masses. Numerically, we arrive at  $(g_\chi^{S} g_t^{S})^{1/2} \in [3.9, 4 \pi]$ and $m_\chi  \in [40,470] \, {\rm GeV}$ $\big($$(g_\chi^{P} g_t^{P})^{1/2} \in [2.2, 4 \pi]$ and $m_\chi  \in [10,580] \, {\rm GeV}$$\big)$. The parameters $\Lambda_{S,P} = 150 \, {\rm GeV}$ and $m_\chi = 50 \, {\rm GeV}$ used in Sec.~\ref{sec:2JetsDM} to simulate the $\Delta \phi_{j_1 j_2}$ distributions have hence been specifically chosen so that  the EFT approach applies and the universe is not over closed. We emphasise that while large regions of parameter space are excluded due to DM overproduction, these bounds can be ameliorated if DM has large annihilation cross sections to other SM particles or (in particular) new hidden sector states. Such additional annihilation channels can reduce the tension between the LHC mono-jet limits and the relic density constraints significantly. 

The preceding discussion should have made clear that the applicability of the LHC mono-jet limits on $\Lambda_{S,P}$ is  limited. This raises the question of whether the jet-jet azimuthal angle difference in $2 j + E_{T, \rm miss}$ remains a good observable to probe the structure of the DM top-quark interactions  also beyond the EFT framework. To answer this question we study a simplified $s$-channel model described by (\ref{eq:LSP}), fixing the relevant parameters to $g^{S,P}_{\chi,t} =1$, $M_{S,P} = 500 \, {\rm GeV}$ and $m_\chi = 200 \, {\rm GeV}$.  Notice that for these parameter choices the DM relic constraints are satisfied. We furthermore verified that our DM models do not lead to an observable signal in existing and future LHC resonance searches in $\bar t t$ (di-jet) final states. Numerically, we find that including the one-loop process $gg \to S,P \to \bar t t$  changes the total $\bar t t$ cross section by~${\cal O} (1\%)$.  A di-jet signal arises in the simplified models (\ref{eq:LSP}) first via the two-loop amplitude $gg \to S,P \to gg$, which renders the contributions of $S,P$ exchange to di-jet production utterly small. 

\begin{figure}[!t]
\includegraphics[width=0.475\textwidth]{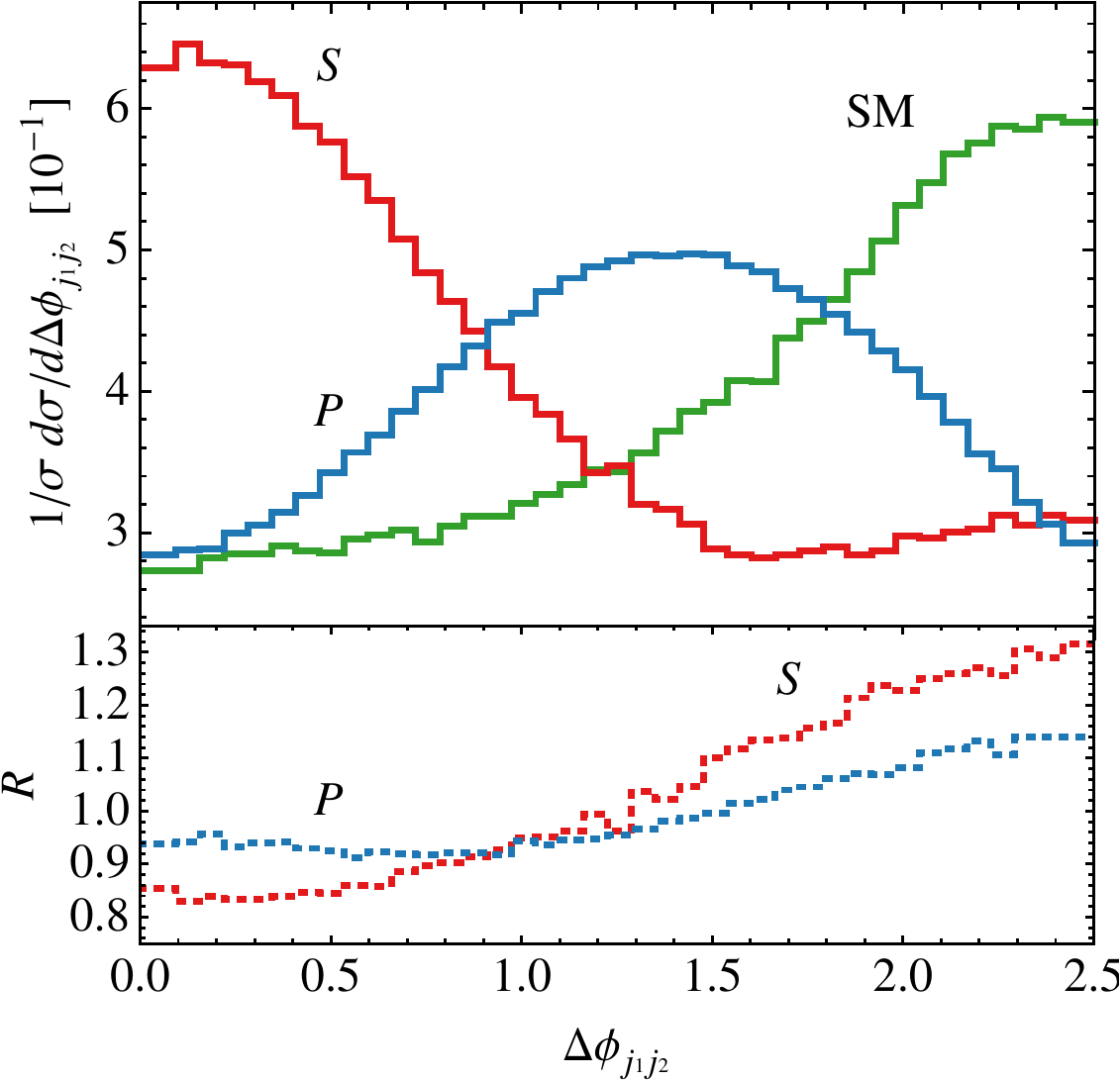}
\vspace{0mm}
\caption{Normalised $\Delta \phi_{j_1j_2}$ distributions arising from a full theory calculation. The used parameters are $g^{S,P}_{\chi,t}  =1$, $M_{S,P} = 500 \, {\rm GeV}$ and $m_\chi = 200 \, {\rm GeV}$ and the shown predictions correspond to our reference cuts. The meaning of the coloured curves is analogue to the one in~Fig.~\ref{Fig2}.}
\label{Fig5}
\end{figure}

The signal strength in $j + E_{T, \rm miss}$ production depends sensitively also on the total widths $\Gamma_{S,P}$ of the mediators~$S,P$. In the case of the scalar mediator, we obtain the following results for the partial decay widths 
\begin{equation} \label{eq:partial}
\begin{split}
\Gamma \hspace{0.25mm} (S \to \bar t t) & = \left ( \frac{m_t}{v} \, g_t^S \right)^2 \frac{3}{8 \pi } \, M_S  \, \bigg ( 1 - \frac{4 m_t^2}{M_S^2} \bigg )^{3/2} \,, \\[2mm]
\Gamma \hspace{0.25mm} (S \to \bar \chi \chi) & = \left ( g_\chi^S \right )^2 \frac{1}{8 \pi } \, M_S  \, \bigg ( 1 - \frac{4 m_\chi^2}{M_S^2} \bigg )^{3/2} \,, \\[2mm]
\Gamma \hspace{0.25mm} (S \to gg) & = \left ( \frac{m_t}{v} \, g_t^S \right)^2  \frac{\alpha_s^2}{2 \pi^3 }  \, \frac{m_t^2}{M_S} \left | F_S \left ( \frac{4 m_t^2}{M_S^2} \right ) \right |^2 \,,
\end{split}
\end{equation}
where 
\begin{equation} \label{eq:FS}
F_S (\tau) = 1 + (1 - \tau) \arctan^2 \left ( \frac{1}{\sqrt{\tau-1}} \right )\,.
\end{equation}
The analog expressions for the pseudo-scalar mediator are obtained from (\ref{eq:partial}) by the replacements $S \to P$ and  $3/2 \to 1/2$ in the exponents, and the relevant form factor reads  
\begin{equation} \label{eq:FP}
F_P (\tau) =  \arctan^2 \left ( \frac{1}{\sqrt{\tau-1}} \right ) \,.
\end{equation}
Using the above values for the couplings and masses, we arrive at $\Gamma_S/M_S = 3.1\%$ and $\Gamma_P/M_P = 6.4\%$, which implies that we are dealing with narrow resonances. The corresponding values of the mono-jet cross sections at the $14 \, {\rm TeV}$ LHC are $\sigma (pp \to j + S (\to \bar \chi \chi)) \simeq 9 \, {\rm fb}$ and  $\sigma (pp \to j + P (\to \bar \chi \chi)) \simeq  25 \, {\rm fb}$, if our standard signal cuts are applied. For the $2 j + E_{T, \rm miss}$  signal cross sections we find instead $\sigma (pp \to 2j + S (\to \bar \chi \chi)) \simeq 5 \, {\rm fb}$  and  $\sigma (pp \to 2j + P (\to \bar \chi \chi)) \simeq 16 \, {\rm fb}$, respectively.\footnote{We recall that the dominant SM backgrounds due to $pp \to j+ Z (\to \bar \nu \nu)$ and $pp \to 2 j+ Z  (\to \bar \nu  \nu)$  have cross sections of $1289 \, {\rm fb}$ and $330 \, {\rm fb}$, respectively.} At the $14 \, {\rm TeV}$~LHC with an integrated luminosity of $300 \, {\rm fb}^{-1}$ one  hence expects to see more than 1000 signal events, which should allow for a measurement  of the $\Delta \phi_{j_1  j_2}$ distribution in the  $2 j + E_{T, \rm miss}$ sample.  

In~Fig.~\ref{Fig5} we show the normalised azimuthal angle distributions corresponding to our explicit DM models.  We see that the strong cosine-like (sine-like) correlation between the two tagging jets in $2 j + E_{T, \rm miss}$ survives in the full theory with resonant scalar (pseudo-scalar) exchange. This shows that, unlike the mono-jet cross section, which depends strongly to the exact model realisation, the normalised $\Delta \phi_{j_1  j_2}$ distribution is rather insensitive to the precise structure of the underlying theory, and therefore provides a unique way to probe the anatomy of possible couplings between DM and top quarks. 

As in the case of the EFT calculations, we also see from the latter figure that the $m_t \to \infty$ approximations of the $\Delta \phi_{j_1  j_2}$ spectra describe the exact results  reasonably well.  We furthermore find that  the heavy top-quark mass limit describes the total  $2 j + E_{T, \rm miss}$ cross sections much better in the full theory than in the EFT framework. Numerically, we obtain for the standard cuts that the ratio of EFT to exact cross sections is around 1.4 for both scalar and pseudo-scalar interactions.  The observed feature is explained by the fact that in the full theory the $\sigma (pp \to 2 j + E_{T, \rm miss})$ cross section is dominated by invariant masses $m_{\bar \chi \chi}$ close to $M_{S,P}$,  while in the EFT calculation  the momentum transfer to the DM pair can be (and is on average) much larger. The quality of the heavy top-quark mass approximation however degrades rapidly with the amount of off-shellness \cite{Haisch:2012kf}, which explains why for the total cross sections the $m_t \to \infty$ limit works fairly well in the case of the simplified model, while it fails badly in the EFT approach. 

\section{Conclusions}
\label{sec:conclusions}

While mono-jet searches  provide already stringent constraints on the pair-production cross sections of DM and may lead to  a future discovery at the LHC, even the observation of an unambiguous $j + E_{T, \rm miss}$ signal will not be enough to determine details of the nature of DM such as the mass of the DM candidate or the structure of its couplings to quarks and gluons. This is due to the fact that while the $p_T$ spectrum of the signal is somewhat harder than that of the background, the enhancement of the high-$p_T$ tail is fairly universal, in the sense that it is independent of the type of interactions that lead to the $j + E_{T, \rm miss}$ events. 

In this letter we have pointed out that some of the limitations of the LHC DM searches can be overcome by studying the jet-jet azimuthal angle difference  in final states with two jets and a large amount of missing transverse energy. We showed in particular that if the  $2 j + E_{T, \rm miss}$ signal arises from Feynman diagrams involving top-quark loops, measurements of the normalised~$\Delta \phi_{j_1  j_2}$ distribution would provide a powerful handle to disentangle whether  the DM top-quark interactions are of scalar or pseudo-scalar type. In contrast to the prediction of the mono-jet cross section that is highly model-dependent, we emphasised that the  strong angular correlation between the two tagging jets is present irrespectively of whether the calculation is performed in an EFT  or in a simplified DM model with scalar and pseudo-scalar exchange in the $s$-channel. This feature combined with the stability of the suggested observable under  QCD corrections, makes  $1/\sigma \, d\sigma (pp \to 2 j + \bar \chi \chi)/d \Delta \phi_{j_1j_2}$ a gold-plated observable to determine the Lorentz structure of the DM top-quark couplings and/or to test the CP properties of the associated mediators.  

The method outlined in our work is more general as, after a DM discovery through a $j + E_{T, \rm miss}$ signal at the LHC, it can in principle be used to tell apart whether DM pair production proceeds dominantly via tree or loop graphs. Only in the latter case, measurements of the azimuthal angle difference  in $2 j + E_{T, \rm miss}$ events can potentially show a strong cosine-like or sine-like modulation, while tree-level exchange of spin-0 and spin-1 mediators will lead to a  distribution with a rather different $\Delta \phi_{j_1j_2}$ dependence. In the case of  discovery, it is hence imperative that ATLAS and CMS study the differential distributions of final states beyond $j + E_{T, \rm miss}$. 

\begin{acknowledgements}
We would like to thank  David~Berge, Felix~Kahlhoefer, Steven~Schramm and Giulia~Zanderighi for useful discussions and their valuable comments on the manuscript. Helpful correspondence with Francisco~Campanario,  Barbara~Jaeger and  Michael~Kubocz  concerning {\tt GGFLO} and  {\tt VBFNLO} is acknowledged.  The research of A.~H. is supported by an STFC Postgraduate Studentship.
\end{acknowledgements}

\end{document}